# Coil geometry with large openings for a HSR3-like stellarator reactor for fast replacement of in-vessel components

V. Queral[1]*, V. Tribaldos[2], J.M. Reynolds[2], I. Fernández[1]

[1.] Laboratorio Nacional de Fusión, CIEMAT, 28040 Madrid, Spain
[2.] Departamento de Física, Universidad Carlos III de Madrid, Leganés, 28911 Madrid, Spain
*E-mail: vicentemanuel.queral@ciemat.es

**Abstract**

Advanced stellarators require convoluted modular coils to produce a plasma with satisfactory performance. Moreover, the number of coils is sometimes high to decrease the modular ripple created by the coils. For reactor stellarators, these requirements imply relatively small ports for in-vessel access and maintenance, i.e. in comparison with tokamaks.

The blankets and divertor modules will have to be replaced periodically (about each 1–4 years depending on the design) due to neutron damage, and also erosion of divertor targets. Blanket modules are activated, thus, all the maintenance operations have to be produced remotely. In order to reduce the shutdown time and cost during component replacement, and to reduce the number, speed and other specifications of the remote maintenance equipment, the number of blanket modules in the reactor should be low and thus, the blanket modules should be large (in relation to the minor and major radius). Nevertheless, the size of the openings between coils limits the maximum size of the blanket and divertor modules, though several potential enhancements have been proposed in the past for stellarators, like straightening the outboard segments of the coils and the movement and/or expansion of certain coils to have wider access.

The present work reports on a coil geometry for the 'Helias Stellarator Reactor' (HSR) of three periods (HSR3) with coils located far from the plasma at the outboard region of the straight-like sector. This feature creates natural wide openings at such regions of the coils, which may be utilized to allow access to large blanket and divertor modules.

## 1. Introduction

Remote maintenance of in-vessel components for stellarators is generally hindered by the usual large number of coils in the reactor, the sometimes large aspect ratio (gives similar outboard an inboard average intercoil distances), and the convoluted shape of the coils, which reduces the free space for large ports. The available size of the ports can be grasped from the figures in i.e. Ref. [1]. Consequences of relatively small ports are: the need of reduction of the maximum size of each blanket module to be manipulated through the port, the subsequent increase in the number of blanket modules per reactor, the increase of the number of feeding pipes and of remote handling operations to extract the full set of blanket modules. The same reasoning is applicable to divertor modules or divertor cassettes to be replaced in the stellarator reactor.

To partially mitigate these difficulties, some concepts have been proposed in the past, like straightening the outboard of certain coils for ARIES-like QA (Quasi-Axisymmetric) configurations [2], or the movement (tilting, displacement or both simultaneously) of enlarged coils for a QI (Quasi-Isodynamic) configuration [3]. Also, the utilization of an enlarged coil at the bean-shape section for certain version of NCSX stellarator [4] was proposed for this experimental non-reactor device.

An option sometimes invoked in tokamaks and stellarators is the expansion of the length of the outboard of the toroidal device. Thus, keeping the number of coils fixed, the coil interspace at the outboard is increased. This works properly for tokamaks, as proposed i.e. in the ARIES-ACT1 tokamak reactor design, see i.e. [5]. However, for stellarators, the procedure might give either positive or poor results. Indeed, the general trend of more convoluted shape of the coils when they are further from the LCFS may hinder the advantages of this possibility for stellarators. One drawback of the expansion of the outboard length, both in tokamaks and stellarators, is the increase in the volume of unused magnetic field, which implies an increase of cost of coils, and a slight increase of the size of the device.

Only the case of three periods is studied. However, the method and coil disposition might be applicable to larger number of periods for HELIAS reactors.

Section 2 defines the methods and Section 3 the results, which includes a final accurate calculation of the neoclassical confinement of the selected coil geometry and the virtual extraction of a large portion of blanket structure.

## 2. Methods

The expansion of the volume between the HSR3 plasma [6] and a winding surface was being sought, partially to try to reduce the size of the reactor, at the expense of extra magnetic field lost in such volume. During such process of expansion, a curious behaviour of certain outboard segments of coils was observed. The work described next aroused from this observation.

HSR3 is a reactor concept of three periods based on the HELIAS magnetic configurations [6,7]. Next, the magnetic configuration of HSR3 is the one coming from the magnetic field produced by the HSR3 coil geometry kindly supplied by the IPP Max Planck (see Acknowledgements). So, it is a configuration that can be generated by coils, for example, the set of coils defined in Ref. [7]. We note, however, that the design of HSR3 is more than 20 years old and that present-day optimizations would produce more attractive physics results. Also, coil geometry compatibility can nowadays be considered as an optimization boundary condition [8] and thus, yield coil geometries much different than the coil geometries designed in former decades.

During such process of expansion of volume, the curvature of the coils located at the centre of the straight-like region of the plasma increased, and contiguous coils approximate to the more central ones. Taking advantage of this property, the size of such outboard region was increased, and larger and free-of-coils openings were obtained.

However, locating the coils at such large distances and creating openings might compromise the initial satisfactory plasma performance of the HSR3 magnetic configuration.

Thus, the search of the existence of coil configurations having large openings and keeping reasonable plasma performance (neoclassical confinement, beta limit, and plasma volume) was started.

Some initial tests on certain parameters were performed, in order to avoid a large set of free parameters. For example, giving other parameters fixed, the number of modes in NESCOIL [9] code was varied, taking either 5 or 6 poloidal and toroidal modes. 6 poloidal modes and 5 toroidal modes reduce the average and maximum errors of the magnetic field on the target LCFS. This exercise was performed for 10 coils per half period only. That number of modes (6 poloidal and 5 toroidal) is kept for all the subsequent analysis in this work.

The winding surface is defined by up to cubic powers of sine and cosine functions of the poloidal and toroidal angles, resulting in 9 combinations. Several tries were performed to select these discrete numbers, so to avoid excessive number of parameters. One of such winding surfaces is the surface containing the coils in Fig. 2.

The number of coils was varied between 5 and 10, giving 6 different values.

During the generation of the winding surface, a parameter that takes the values 0.045, 0.05 and 0.055 is considered, which mainly varies the inboard of the quasi-straight regions of the plasma (3 different values).

It results 162 cases (9 x 6 x 3). For each case, the coils on the winding surface are obtained, then the resulting magnetic field is obtained by means of Biot Savart law, and the resulting LCFS Fourier coefficients are obtained with the CASTELL code [3,10].

The approximate neoclassical confinement of each case was computed by CASTELL code directly from particle orbits at three minor radius $\rho$. This approximate average neoclassical confinement was utilized as a factor for the selection of the 'best' case. The accurate calculation of neoclassical confinement for the selected case was performed by the MOCA code [11], see Section 3.1.

Mercier and ballooning plasma stability was computed for each case by the Variational Moments Equilibrium Code (VMEC) [12] and the Code for Ballooning Rapid Analysis (COBRA) [13]. Each case is associated with six different beta VMEC runs (for $\beta(\%) \approx$ 0.9, 1.3, 1.8, 2.2, 2.7 and 3.1). From the VMEC output of each run, the Mercier stability is obtained on each radial position. The result is a parameter scan in the radial direction (40 radial positions) and beta (6 runs). Counting the total number of stable occurrences in the scan against the total number of occurrences (40x6=240), a *proportion of stable occurrences* can be obtained. This parameter is used as a factor for the selection of the 'best' case of the 162 cases. The same procedure is followed for the Ballooning stability, coming the *stable occurrences* from six COBRA runs which utilize the same six VMEC-outputs.

Also, the approximate plasma volume created by the coils was computed by CASTELL. This has to be considered in this type of searches since some coil configurations (some cases) may give rather satisfactory neoclassical confinement and plasma stability at expense of too reduced plasma volume, which is undesirable.

A potential issue while trying to obtain larger openings at the outboard may be the concentration of current density at certain areas. Considering coils on typical winding surfaces (not in the free space), any region that does not contain, or contain low, current density has to be compensated by higher current density in other area. The location of the outboard winding surface further from the plasma (larger length and surface) partially compensates the current increase, but, still, it may be an issue for the real construction of the coils.

## 3. Results

Among the 162 cases, there are several with high neoclassical confinement similar to the one corresponding to the original from the HSR3 coils. Among this, some have beta limit lower but approaching to that of the original HSR3 coils. Also, a plasma volume similar to the original HSR3 configuration is taken into consideration to select one notable coil geometry on a particular winding surface.

A coil geometry, called C161, is selected among the others due to a good balance of the three parameters for the resultant plasma.

Certainly, the inboard part of the coils is rather convoluted in all the cases. No effort has been performed to smooth such inboard part of the coils, since here the focus is the outboard. Likely, nearing the winding surface at the inboard region would reduce the current concentration and convolution of coils at the inboard.

Fig. 1 shows the toroidal cuts of the LCFSs of the plasma coming from the original HSR3 coils (blue), and the resultant LCFS from the C161 coil geometry (red). The toroidal cuts for the C161 case come from the Fourier coefficients of a surface adjusted to the outermost closed particle-orbit from the simulation of the particle in the C161 coil geometry. An outermost non-low-rational orbit/surface is selected in order to have a good coverage of the LCFS.

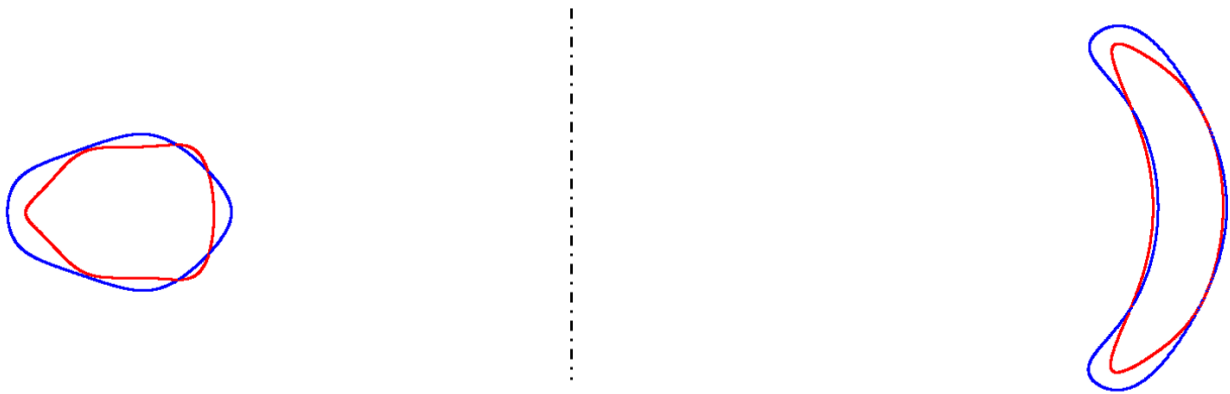

**Fig. 1**. Toroidal cuts of the LCFSs of the plasma coming from the original HSR3 coils (blue), and the resultant LCFS from the C161 coil geometry (red).

Section 3.1 reports on the accurate calculation of mono-energetic transport coefficients $D_{11}$ for HSR3 and C161 configurations, for $\beta = 0$ and $\beta = 5\%$. The neoclassical confinement resulted less than two-fold higher for C161 than for the original HSR3.

Beta limit is also reduced in C161 compared to HSR3. According to the Methods in Section 2, Mercier stability is 48% for C161 and 74% for HSR3. Also according to the Methods, ballooning stability is calculated 62% for C161 and 89% for HSR3. It results 30-35% reduction in beta limit for the coil geometry

C161. In principle, it is considered satisfactory, in view of the important advantages and much lower cost of remote maintenance of the blanket modules, and lower related shutdown costs.

Fig. 2 shows the C161 coil geometry and the 3D representation of the resultant LCFS generated from such coils. Fig. 2 is a quasi-plan view of the reactor, which has a slight rotation with respect the 'X-Axis' (Fig. 2-a) starting from a pure plan view. Also, Fig. 2 shows the comparative size of the created openings (defined by the $r_v$ $t_v$ dimensions) with the size of the plasma at such region.

The dimensions $r_v$ $t_v$ mean:

$r_v$ : Approximate radial space for vertical extraction of components (thickness of the coil winding packs and casings, and the thickness of the vacuum vessel are not included).

$t_v$ : Approximate toroidal space for vertical extraction of components (same comment as above).

Fig. 2-b shows that:

$r_v$ is similar to twice the plasma minor radius at such region, which is the quasi-straight region of the plasma.

$t_v$ is similar to the length of the quasi-straight region of the plasma.

This gives a wide space for replacement (extraction) of neutron damaged or failed blanket modules and the introduction of new blanket modules.

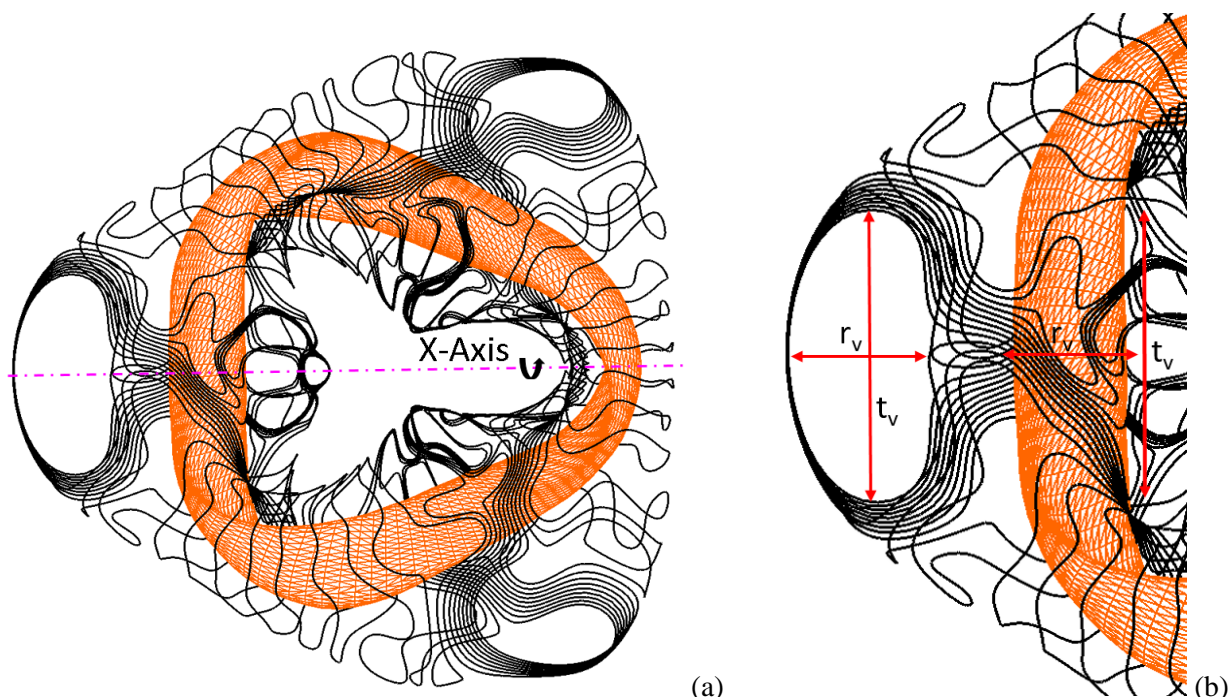

**Fig. 2**. **a)** C161 coil geometry (black) and resulting plasma from such coils (orange) seen from a point of view having a slight rotation with respect the 'X-Axis'. The plasma corresponds to the red poloidal cuts in Fig. 1. **b)** Detail of figure (a) comparing the size of the plasma (and so, of the blanket modules) and the size of the openings in radial and toroidal direction for this particular angle of vision.

Blanket modules, fully encircling the plasma, located at the quasi-straight region of the plasma (Fig. 2-b, length $t_v$) of length half of the quasi-straight region ($t_v/2$) may be extracted/introduced, as shown in Section 3.2. If $r_v$ were insufficient to allow replacement of a full portion of blanket toroid (blanket module) (see Fig. 5), it could be divided in two poloidal parts. Section 3.2 studies in more detail rotations/translations of such possible large blanket module, to take advantage of the full available opening space, which depends on the line of sight, i.e. see Fig. 3-yellow opening.

Certainly, space for the winding packs and coils cases has to be allocated. However, the coils are considered filamentary in this first approximation. At the location of the concentration of coils (region C, Fig. 3), some of the 8 coils have to be displaced outwardly and others inwardly (and toroidally), in order to create space for the winding pack and coil casings. This will reduce somewhat the free space for remote maintenance operations. Also, the thickness of the walls of the ports and the thickness of the vacuum vessel have to be considered.

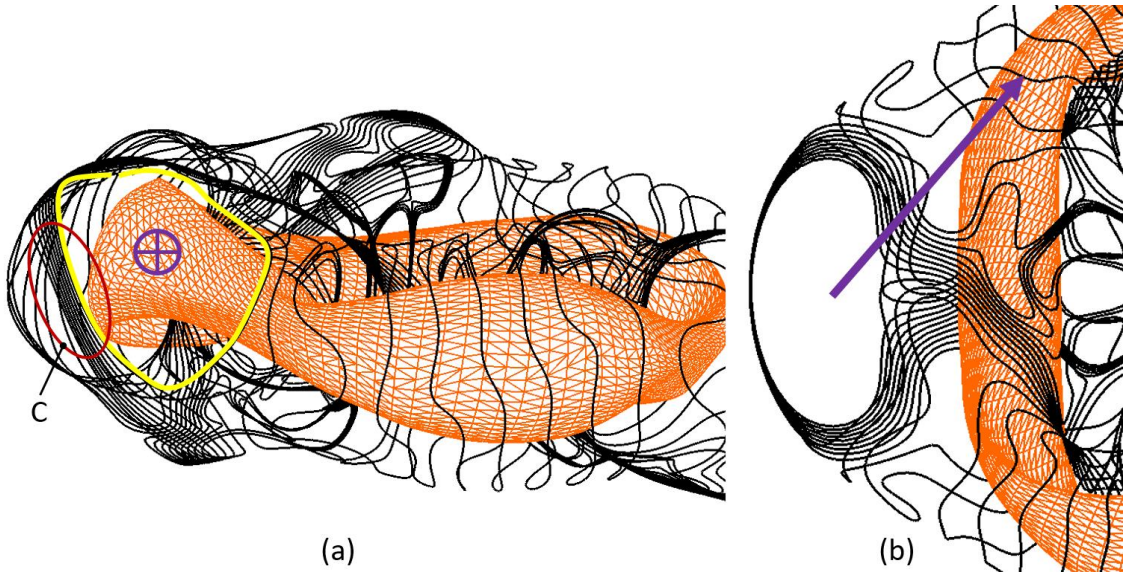

**Fig. 3**. The free space for access in the direction of the violet arrow (b) is shown in yellow in (a), and a violet cross corresponding to the arrow is represented (a). This direction may allow extracting blanket modules located at the curved plasma sectors following almost a linear movement. Blanket modules located at the quasi-straight plasma sector will require rotations plus translations.

### 3.1 Accurate calculation of neoclassical confinement for the selected case

The estimation of the neoclassical confinement of the plasma created by each coil geometry (162 cases) was performed by the CASTELL code, which implements a fast neoclassical transport estimation. Thus, a final calculation of the selected C161 case is performed with the MOCA code [11, 14].

The neoclassical confinement of the original HSR3 configuration is compared with the C161 magnetic configuration (coming from the coils). Fig. 4 shows the normalized mono-energetic transport coefficient $D_{11}$ (as defined in [14]) for HSR3 and C161 magnetic configurations, for $\beta = 0$ and $\beta = 5\%$. The plot has been calculated for the flux surface at $\rho = 0.5$.

For $\beta = 0$, $D_{11}$ is about two-fold higher for C161 than for the original HSR3, Fig. 4. For some collisionalities, the reduction in neoclassical confinement is lower than two.

For $\beta = 5\%$, $D_{11}$ is only slightly higher for C161 than for the original HSR3.

This result appears satisfactory, since even larger reductions in neoclassical confinement may be acceptable, as observed from Ref. [15], where the neoclassical confinement of W7-X is compared with the much lower one in LHD. Future optimization might find a coil geometry having even lower $D_{11}$ than the case C161 and still retaining large ports.

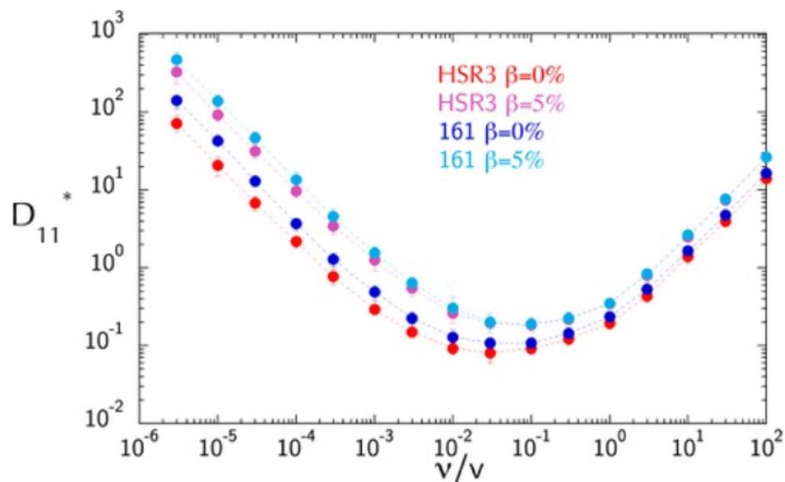

**Fig. 4**. Mono-energetic transport coefficient $D_{11}$ for HSR3 and C161 configurations for $\beta = 0$ and $\beta = 5\%$ at $\rho = 0.5$

## 3.2 Application to the approximate maximum size of blanket modules

Filamentary winding packs and coil casings are assumed for this first approximation. The space for the vacuum vessel, thermal shields and gaps are also limited in certain curved sectors. This implies somewhat smaller maximum size of the possible blanket modules.

Fig. 5 shows a continuous blanket structure of 0.99 m thickness around the plasma. The distance from the LCFS to the inner surface of the blanket structure (first wall) is 0.2 m. This thickness includes a breeding layer and radiation shielding. Part of the radiation shielding would be produced by the vacuum vessel, though there is little space for the vacuum vessel at some areas at the curved sector. It will be improved in the future.

The size represented in Fig. 5 corresponds to the size of the HSR3 magnetic configuration and plasma [6, 7], having major radius $R$~15m. Certainly, the blanket plus shielding thickness is, to some extent, fixed and thus, a large size of the reactor has to be considered to allow radial space for breeding and shielding.

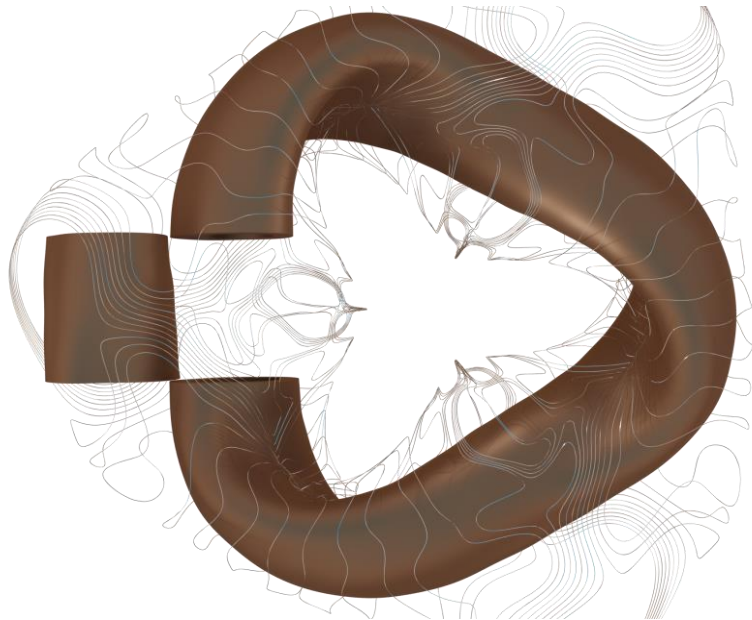

**Fig. 5**. Plan view of a monolithic blanket toroid, the coil geometry and a 'cut' portion of blanket toroid (length 9 m), which is moved radially. The portion of blanket represents a possible large blanket module still to be developed.

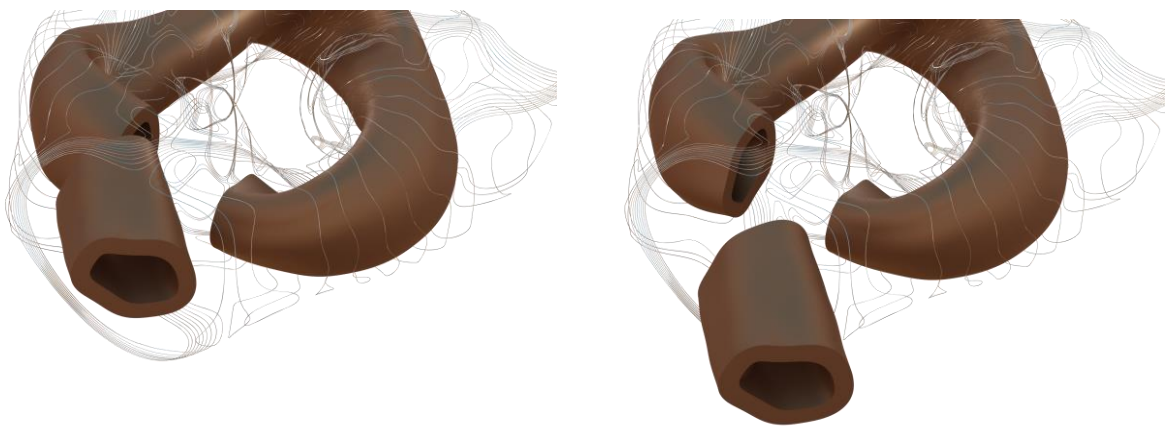

**Fig. 6**. Perspective view of Fig. 5 with the portion of blanket rotated and translated for subsequent extraction through the large port (left), and the same perspective view with the portion of blanket already extracted (right).

The minimum number of blanket modules cannot be deduced from this simplified study. Nevertheless, 30 modules might be a guess (conservatively it is assumed that the portion of blanket module in Fig. 6 is split in two parts to allow extraction). Further studies are required to know the maximum size of modules at the curved sectors.

## 4. Conclusions and future work

Large openings in a coil geometry for a HELIAS stellarator reactor of three periods (HSR3) has been obtained, and a method to find this type of coil geometries has been developed and applied. Although the coil system of the HELIAS reactor configuration HSR3 is an old one, the methods developed here are general enough to be applied to a diversity of coil designs and magnetic configurations.

The coils are located far from the plasma at the outboard region of the straight-like sector of the plasma, giving rise to natural openings at the coil structure. Thus, coils are concentrated in certain areas so to allow wide openings in other areas (a basic geometric property). Concentration of coils and currents is produced at the outboard region of the straight-like plasma sector. Still, the large outboard length around the torus slightly decreases the current concentration. Space for the winding packs and coil casing is necessary, which will somewhat reduce the free space at the opening. The inboard part of the coils has not been refined and, likely, unnecessary current concentration is occurring. The neoclassical confinement estimated for Case 161 (C161) by the MOCA code is better than half the original one for HSR3 magnetic configuration. Beta limit is 30-35% lower than for the original HSR3 configuration. This coil geometry losses abundant magnetic field/energy in the large volume at the outboard straight-section.

Extraction/introduction of blanket modules as long as 9 m and having a full poloidal turn of the blanket (or split in two poloidal segments) appears feasible, which represents an important improvement with respect the possibilities of common ports for HELIAS stellarators, i.e. see [1]. This feature may be relevant to try to much reduce the shutdown time and cost during remote maintenance, and to reduce the cost of the remote handling equipment. Nevertheless, coil/current concentration, geometrically complex coils, and loss of magnetic field/energy are important drawbacks of this concept. Also, the applicability to other magnetic configurations is not assured.

The study of the potential application of the method to higher number of periods (i.e. 4, 5 periods) for HELIAS-like magnetic configurations and for possible QI configurations remain for a future work.

**Acknowledgement**
The authors are grateful to the Stellarator Theory Department of the Max-Planck-Institute for Plasma Physics (Greifswald, Germany) for providing and granting the use of the HSR3 configuration for this publication, Raúl Sánchez (Univ. CIII de Madrid) for COBRA assistance, and Edilberto Sánchez (CIEMAT) for help in parallel computing.

This work was partly funded by the 'Agencia Estatal de Investigación' (AEI), the 'Ministry of Science, Innovation and Universities', and by the 'Fondo Europeo de Desarrollo Regional' (FEDER), under the Grant PID2021-123616OB-I00, for the project "Study of improved stellarator assemblies consistent with proper in-vessel components for viable high-field stellarator reactors"; by the same institutions under the previous grant RTI2018-098356-B-I00, project 'Study of additive and advanced manufacturing for high magnetic field fusion devices of stellarator type'; and by the same institutions under the grant PID2022-137869OB-I00. Calculations were done in Uranus, a supercomputer cluster located at Universidad Carlos III de Madrid and funded jointly by EU-FEDER and the Spanish Government via Project No. UNC313-4E-2361, ENE2009-12213-C03-03, ENE2012-33219 and ENE2015-6826.